\documentclass[prl,twocolumn]{revtex4}
\usepackage{graphicx}
\usepackage{dcolumn}
\usepackage{color}
\usepackage{latexsym}
\usepackage{bm}
\begin{document}
\title{Hawking Radiation and Covariant Anomalies}
\author{Rabin Banerjee}
\email{rabin@bose.res.in}
\author{Shailesh Kulkarni}
\email{shailesh@bose.res.in}
\affiliation{
Satyendra Nath Bose National Centre for Basic Sciences, Block-JD, 
Sector-III, Salt Lake, Kolkata 700 098, India
}

\begin{abstract}
 Generalising the method of Wilczek and collaborators we provide
 a derivation of Hawking radiation from charged black holes using 
 only covariant gauge and gravitational anomalies. The reliability 
 and universality of the anomaly  cancellation approach to Hawking 
 radiation is also discussed.   
\end{abstract}
\maketitle
{\it Introduction:-}\\
 Hawking radiation is an important quantum effect
 in black hole physics. Specifically, it arises in the background spacetime
 with event horizons. The radiation has a spectrum with Planck distribution
 giving the black holes one of its thermodynamic properties that make it
 consistent with the rest of physics. Hawking's original result 
\cite{Hawking} has since been rederived in
 different ways thereby reinforcing the conclusion to a certain extent.
 However, the fact that no one derivation is truly clinching has led to
 open problems leading to alternative approaches with fresh insights.

        An anomaly in quantum field theory is a breakdown of some
 classical symmetry due to the process of quantization (for reviews, see
 \cite{rabin1,bertlmann,Fujikawa}). Specifically, for instance, a gauge
 anomaly is an anomaly in gauge symmetry, taking the form of nonconservation
 of the gauge current. Such  anomalies characterise a theoretical 
 inconsistency, leading to problems with the probabilistic interpretation
 of quantum mechanics. The cancellation of gauge anomalies gives strong
 constraints on model building. Likewise, a gravitational anomaly 
\cite{Witten,Bardeen} is an anomaly in general covariance, taking the
 form of nonconservation of the energy-momentum tensor. There are other
 types of anomalies but here we shall be concerned with only gauge 
 and gravitational anomalies. The simplest case for these anomalies
 which is also relevant for the present analysis, occurs for $1+1$ dimensional
 chiral fields.        

   Long back Christiansen and Fulling \cite{Fulling} reproduced Hawking's 
 result by exploiting the trace anomaly in the energy momentum tensor of
 quantum fields in a Schwarzchild black hole background. The use of
 anomalies, though in a different form, has been powerfully resurrected
 recently by Robinson and Wilczek \cite{Robwilczek}. They observed that
 effective field theories become two dimensional and chiral near the
 event horizon of a Schwarzchild black hole. This leads to a two
 dimensional gravitational anomaly. The existence of energy flux of 
 Hawking's radiation is necessary to cancel this anomaly. The method of
 \cite{Robwilczek} was soon extended to charged black holes \cite{Isowilczek}
 by using the gauge anomaly in addition to the gravitational anomaly. 
Further advances and applications of this approach may be found in a
 host of papers \cite{Isoumtwilczek,Muratasoda1,Vagenas,Setare,chen,Morita,Jiang,Wu,Cai,Zhang,Shin,Peng,Jiang2,He,Murata2,Isomorita}, including a recent
 review \cite{das}. 

     The approach of \cite{Robwilczek,Isowilczek} is based on the fact that 
 a two dimensional chiral (gauge and/or gravity) theory is anomalous. Such theories admit two types
 of anomalous currents and energy momentum tensors; the consistent
 and the covariant \cite{rabin1,bertlmann,Fujikawa}. The covariant divergence of these  currents and energy-momentum tensors yields either the 
consistent or the covariant form of the gauge and  gravitational anomaly, 
respectively \cite{Witten,Bardeen,rabin2,rabin1,bertlmann,Kolprath,Fujikawa}.
 The consistent current and anomaly satisfy the Wess Zumino consistency
 condition but do not transform covariantly under a gauge transformation.
  Expressions for the covariant current and anomaly, on the contrary,
 transform covariantly under gauge transformation but do not satisfy
  the Wess Zumino condition. Similar conclusions also hold for the
 gravitational case, except that currents are now replaced by
 energy-momentum tensors and gauge transformations by general coordinate
 transformations. In\cite{Robwilczek,Isowilczek} the charge and the energy momentum flux  of the Hawking radiation is obtained by a cancellation of 
the consistent  anomaly. However the boundary condition necessary to 
fix the parameters 
 are obtained from a vanishing of the covariant current at the event
 horizon. 

    In this paper we generalise the method of \cite{Robwilczek,Isowilczek} by presenting a unified description totally in terms of covariant
  expressions. This discussion is specifically done for Hawking radiation
 from charged black holes. The charge flux is determined by a cancellation 
 of the covariant gauge anomaly while the energy momentum flux
 is fixed by cancellation of the covariant gravitational anomaly. These
 are the only inputs. Also, we show that the analysis of 
 \cite{Robwilczek,Isowilczek} is resilient and the results are unaffected
 by taking more general expressions for the consistent anomaly which
 occur due to peculiarities of two dimensional spacetime. \\

 {\it General discussion on covariant and consistent anomalies:-}\\
 Here we  briefly summarise some results on anomalies highlighting the
 peculiarities of two dimensional spacetime. First, the consistent gauge
 anomaly as taken in \cite{Robwilczek, Isowilczek} is considered,
\begin{equation}
\nabla_{\mu}J^{\mu} = \pm \frac{e^2}{4\pi} \bar \epsilon^{\rho\sigma}\partial_{\rho}A_{\sigma} = \pm \frac{e^2}{4\pi \sqrt{-g}} \epsilon^{\rho\sigma}\partial_{\rho}A_{\sigma} \label{1}
\end{equation}
where $+(-)$ corresponds to left(right)-handed fields, respectively.
 Here $g_{\mu\nu}$ is the two dimensional $(r-t)$ part of the complete Reissner-Nordstrom metric given by 
\cite{Robwilczek, Isowilczek}
\begin{equation}
ds^2 = f(r)dt^2 -\frac{1}{f(r)}dr^2 - r^2 d\Omega^{2}_{(d-2)}. \label{2}  
\end{equation} 
so that $-g = -det g_{\mu\nu} = 1$ and $d\Omega^{2}_{(d-2)} $ is the 
line element on the $(d-2)$ sphere.   
The gauge potential is defined as $A = -\frac{Q}{r}dt$.

    Now a word regarding our conventions. As is evident from (\ref{1}) the
 antisymmetric tensor $\bar\epsilon^{\rho\sigma}$ differs from its
 numerical counterpart $\epsilon^{\rho\sigma}$ ($\epsilon^{01} = -\epsilon_{01} = 1$) by the factor $\sqrt{-g}$. Since here $\sqrt{-g} =1$, the two get
 identified. Henceforth we shall always use $\epsilon^{\rho\sigma}$,
 omitting the $\sqrt{-g}$ factor.   

  The current $J^{\mu}$ in (\ref{1}) is called the consistent current and 
 satisfies the Wess-Zumino consistency condition. Effectively this
 means that the following integrability condition holds \cite{rabin1,rabin2};
\begin{equation}
\frac{\delta J^{\mu}(x)}{\delta A_{\nu}(y)} = \frac{\delta J^{\nu}(y)}{\delta A_{\mu}(x)}. \label{3}
\end{equation}
The covariant divergence of the consistent current yields the consistent 
anomaly. The structure appearing in (\ref{1}) is the minimal form, since only
 odd parity terms occur.  However it is possible that normal parity 
 terms appear in (\ref{1}). Indeed, as we now argue, such a term is a natural consequence of two dimensional
 properties.

    To fix our notions, consider the interaction Lagrangian for a chiral
 field $\psi$ in the presence of an external gauge potential $A^{\mu}$
 in $1+1$ dimensions,
 \begin{equation}
\mathcal{L}_{I} = \bar\psi \left(\frac{1 \pm \gamma_{5}}{2}\right)\gamma_{\mu}
A^{\mu}\psi.\label{4}    
\end{equation}
Using the property of two dimensional $\gamma$- matrices,
\begin{equation}
\gamma_{5}\gamma^{\mu} = - \epsilon^{\mu\nu}\gamma_{\nu}, \label{5}
\end{equation} 
it is found that $A_{\mu}$ couples as a chiral combination $(g^{\mu\nu} \pm \epsilon^{\mu\nu})A_{\nu}$. Note that the usual flat space identity (\ref{5})
 holds due to the specific structure of the two dimensional metric. 
 Hence the expression for the anomaly in (\ref{1}) generalises to,
\begin{equation}
\nabla_{\mu}\bar J^{\mu} = \partial_{\mu}\bar J^{\mu} = \pm \frac{e^2}{4\pi} \partial_{\alpha}[(\epsilon^{\alpha\beta} \pm g^{\alpha\beta})A_{\beta}].\label{6}
\end{equation}
This is a non-minimal form for the consistent anomaly dictated by the symmetry
 of the Lagrangian, and has appeared earlier in the literature \cite{Bardeen}. 
 It is clear that if $J^{\mu}$ is a consistent current
 then $\bar J^{\mu}$, which is given by, 
\begin{equation}
\bar J^{\mu} = J^{\mu} + \frac{e^2}{4\pi} A^{\mu} \label{7} 
\end{equation}
is also a consistent current since the extra piece satisfies the 
 integrability condition (\ref{3}).

  It is possible to modify the new consistent current (\ref{7}), by adding
 a local counterterm, such that it becomes covariant,
\begin{equation}
\tilde{J^{\mu}} = \bar J^{\mu} \mp \frac{e^2}{4\pi } A_{\alpha}(\epsilon^{\alpha\mu} \pm g^{\alpha\mu}). \label{8}
\end{equation}
The current $\tilde J^{\mu}$ yields the gauge covariant anomaly,
\begin{equation}
\nabla_{\mu}\tilde{J^{\mu}} = \pm\frac{e^2}{4\pi}\epsilon^{\alpha\beta}F_{\alpha\beta}.\label{9} 
\end{equation}
Note that the covariant current (\ref{8}) does not satisfy the Wess-Zumino
 consistency condition since the counterterm violates the integrability
 condition (\ref{3}). Moreover the gauge covariant anomaly (\ref{9}) has
 a unique form dictated by the gauge transformation properties. This
 is contrary to the consistent anomaly which may have a minimal (\ref{1})
or non-minimal (\ref{6}) structure.
 
   Now we will concentrate our attention 
 on the gravity sector. If we omit the ingoing modes the energy momentum 
tensor near the horizon will not conserve, while
 there is no difficulty in the region outside the horizon. The analysis
  \cite{Robwilczek,Isowilczek} for obtaining the flow of energy momentum tensor   was done by using the minimal form  of the consistent $d=2$ anomaly
 \cite{Witten,Bardeen,bertlmann,Kolprath}, for right handed fields,
\begin{equation}
\nabla_{\mu}T^{\mu}_{\nu} = \frac{1}{96\pi}\epsilon^{\beta\delta}
\partial_{\delta}\partial_{\alpha}\Gamma^{\alpha}_{\nu\beta},\label{24}
\end{equation}
 Here we consider the general form for $d=2$ consistent gravitational anomaly.
  It is worthwhile to point out that the consistent gravitational anomaly and 
the consistent gauge anomaly are analogous satisfying similar consistency
 conditions. This is easily observed here by comparing (\ref{24}) with
 (\ref{1}) where the affine connection plays the role of the gauge potential.
 We therefore omit the details and write the generalised anomaly by an
 inspection of (\ref{6}) on how to include the normal parity
 term. The result is,
\begin{equation}
\nabla_{\mu}\bar T^{\mu}_{\nu} = \frac{1}{96\pi}\partial_{\delta}\partial_{\alpha}\left[(\epsilon^{\beta\delta} + g^{\beta\delta})\Gamma^{\alpha}_{\nu\beta}\right] = \mathcal{A_{\nu}}.\label{25} 
\end{equation}

The covariant energy momentum tensor, on the other hand, has the 
divergence anomaly,
\begin{equation}
\nabla_{\mu}\tilde T^{\mu}_{\nu} = \frac{1}{96\pi}\epsilon_{\nu\mu}\partial^{\mu}R = \tilde \mathcal{A}_{\nu}.\label{35}
\end{equation}
This is called the covariant anomaly as distinct from the consistent anomaly
 (\ref{24}).\\
      
{\it Covariant gauge anomaly and charge flux:-} \\
The current is conserved  outside the horizon so that $\nabla_{\mu}\tilde J^{\mu}_{(o)} = \partial_{\mu}\tilde J^{\mu}_{(o)}= \partial_{r}\tilde J^{r}_{(o)} =0$. Near the horizon there are only outgoing (right-handed) fields and the current becomes (covariantly) anomalous (\ref{9}),
\begin{equation}
\partial_{r}\tilde J^{r}_{(H)} = \frac{e^2}{2\pi} F_{rt} = \frac{e^2}{2\pi} \partial_{r}A_{t}. \label{gaugeanomaly}  
\end{equation}
The solution in the different regions is given by,
\begin{eqnarray}
\tilde J^{r}_{(o)} = c_{o},\label{current1}\\
\tilde J^{r}_{(H)} = c_{H} + \frac{e^2}{2\pi}[A_{t}(r) - A_{t}(r_{+})],\label{current} 
\end{eqnarray} 
 where $c_{o}$ and $c_{H}$ are integration constants. 

   The current is now written as a sum of two contributions from the 
 two regions, $\tilde J^{\mu}= \tilde J^{\mu}_{(o)}\Theta(r-r_{+} -\epsilon) + \tilde J^{\mu}_{(H)}H$, where $H = 1- \Theta(r-r_{+} -\epsilon)$. Then by using the conservation 
 equations, the Ward identity becomes,
\begin{eqnarray}
\partial_{\mu} \tilde J^{\mu} = \partial_{r}\tilde J^{r} = \partial_{r}\left( \frac{e^2}{2\pi}A_{t}H\right)  \ + \nonumber\\
  \delta(r - r_{+} - \epsilon)(\tilde J^{r}_{(o)} - \tilde J^{r}_{(H)} + 
 \frac{e^2}{2\pi}A_{t}). \label{gaugeward}
\end{eqnarray} 
To make the current anomaly free the first term must be canceled by quantum
effects of the classically insignificant ingoing modes. This is the Wess-Zumino
term induced by these modes near the horizon. Effectively it implies a
 redefinition of the current as $\tilde J'^{r} = (\tilde J^{r} - \frac{e^2}{2\pi}A_{t}H)$ 
which is anomaly free provided the coefficient of the delta function 
vanishes, leading to the condition,
\begin{equation}
c_{o} = c_{H} -\frac{e^2}{2\pi}A_{t}(r_{+}). \label{constants}
\end{equation}
The coefficient $c_{H}$ is fixed by requiring the vanishing of the 
 covariant current at the horizon. This yields $c_{H} =0$ from
 (\ref{current}). Hence the value of the charge flux is given by,
 \begin{equation}
c_{o} = -\frac{e^{2}}{2\pi}A_{t}(r_{+}) =  \frac{e^{2}Q}{2\pi r_{+}}.
 \label{chargeflux}
\end{equation}
 This is precisely the current flow of the Hawking blackbody radiation
 with a chemical potential \cite{Isowilczek}.\\

{\it Covariant gravitational anomaly and energy-momentum flux:-}\\
 In the presence of a charged field the classical energy-momentum tensor is no
 longer conserved but gives rise to the Lorentz force law, 
 $\nabla_{\mu}\tilde T^{\mu}_{\nu} = F_{\mu\nu}\tilde J^{\mu}$. The corresponding
 anomalous Ward identity for covariantly regularised quantities is then
 given by,
\begin{equation}
\nabla_{\mu}\tilde T^{\mu}_{\nu} = F_{\mu\nu}\tilde J^{\mu} + \tilde\mathcal{A}_{\nu},\label{gravityward}
\end{equation} 
where $\tilde\mathcal{A}_{\nu}$ is the covariant gravitation anomaly (\ref{35}).
 Since the current $\tilde J^{\mu}$ itself is anomalous one might
 envisage the possibility of an additional term in (\ref{gravityward}) 
 proportional to the gauge anomaly. Indeed
 this happens in the Ward identity for consistently regularised objects
 \cite{Isowilczek}. Such a term is ruled out here because there is no such covariant piece with the correct dimensions, having one free index.

   For the metric (\ref{2}) the covariant anomaly is purely time-like 
 $(\tilde\mathcal{A}_r =0)$ while, 
\begin{eqnarray}
\tilde\mathcal{A}_t = \partial_{r}\tilde N^{r}_{t}; \  \tilde N^{r}_{t} = \frac{[ff'' - \frac{(f')^2}{2}]}{96\pi}. \label{A} 
\end{eqnarray}
 Next, the Ward identity is solved for the $\nu= t$ component. In the exterior 
 region there is no anomaly and the Ward identity reads,
\begin{equation}
\partial_{r}\tilde T^{r}_{t(o)} = F_{rt}\tilde J^{r}_{(o)}. 
\end{equation}
Using (\ref{current1})this is solved as
\begin{equation}
\tilde T^{r}_{t(o)} = a_{o} + c_{o}A_{t}(r), \label{emoutside} 
\end{equation}
where $a_{o}$ is an integration constant. Near the horizon the anomalous Ward 
 identity, obtained from (\ref{gravityward}), reads
\begin{equation}
\partial_{r}\tilde T^{r}_{t(H)} = F_{rt}\tilde J^{r}_{H} + \partial_{r}\tilde N^{r}_{t},
\end{equation}  
Using $\tilde J^{r}_{(H)}$ from (\ref{current}) yields the solution
\begin{equation}
\tilde T^{r}_{t(H)} = a_{H} + \int^{r}_{r_{+}} dr \partial_{r}\left[c_o A_{t}+
\frac{e^2}{4\pi}A_{t}^2 + \tilde N^{r}_{t}\right]. \label{emhorizon}
\end{equation} 
Writing the energy-momentum tensor as a sum of two combinations 
$\tilde T^{\mu}_{\nu} = \tilde T^{\mu}_{\nu(o)} \Theta(r-r_+ -\epsilon) +  \tilde T^{\mu}_{\nu(H)}H$
 we find
\begin{eqnarray}
\nabla_{\mu}\tilde T^{\mu}_{t} = \partial_{r}\tilde T^{r}_{t}=c_o \partial_{r}A_{t}(r) +\partial_{r}\left[(\frac{e^2}{4\pi}A_{t}^2 + \tilde N^{r}_{t})H\right] +  \nonumber\\
(\tilde T^{r}_{t(o)} - \tilde T^{r}_{t(H)} + \frac{e^2}{4\pi}A_{t}^2 + \tilde N^{r}_{t}) \delta(r-r_{+} -\epsilon).
\end{eqnarray} 
The first term is a classical effect coming from the Lorentz force. 
 The second term has to be canceled by the quantum effect of the incoming modes.  As before, it implies the existence of a Wess-Zumino term modifying
 the energy-momentum tensor as $\tilde T'^{\mu}_{t} =\tilde  T^{\mu}_{t} - \left[(\frac{e^2}{4\pi}A_{t}^2 + \tilde N^{r}_{t})H\right]$ which is anomaly free provided
 the coefficient of the
 last term vanishes. This yields the condition, 
\begin{equation}
a_{o} = a_{H} + \frac{e^2}{4\pi}A_{t}^2 (r_{+}) - \tilde N^{r}_{t}(r_{+}).
\end{equation}
 where the integration constant $a_H$ is fixed by requiring that the 
covariant energy momentum tensor vanishes at the horizon. From (\ref{emhorizon}) this gives $a_{H} = 0$. Hence the total flux of the energy momentum tensor
 is given by
\begin{equation}
a_o = \frac{e^2}{4\pi}A_{t}^2 (r_{+}) - \tilde N^{r}_{t}(r_{+}). \label{emflux}
\end{equation} 
 Since $f(r_{+}) = 0$ we find from (\ref{A}) that $\tilde N^{r}_{t}(r_{+}) = -\frac{(f')^2|_{r_+}}{192\pi}$. Using the surface gravity of the black hole $\kappa = \frac{2\pi}{\beta} = \frac{(f')|_{r_+}}{2}$, the final
 result is expressed in terms of the inverse temperature $\beta$ as
\begin{equation}
a_o = \frac{e^2 Q^2}{4\pi r_{+}^2} + \frac{\pi}{12\beta^2}.
\end{equation}
This is just the energy flux from blackbody radiation with a chemical potential
 \cite{Isowilczek}.\\

{\it Generalised consistent anomaly and flux:-}\\
 Here we show that the conclusions of \cite{Robwilczek,Isowilczek} remain unaffected by taking the general form
 of the consistent anomaly (\ref{6}, \ref{25}). Instead of repeating their
 analysis we just point out the reasons for this robustness.

   For static configuration and for the specific choice of the potential 
    $(A_{r} = 0)$, it is clear that the normal parity term in (\ref{6})
 vanishes. Likewise the normal parity term in the counterterm (\ref{8})
 also vanishes since only the $\mu=r$ component in $J^{\mu}$ is relevant.
 Hence , effectively the same structures of the consistent (gauge) anomaly
 and the counterterm relating the consistent and covariant currents,
 as used in \cite{Isowilczek}, are valid. Since these were the two
 basic inputs the results concerning the charge flux associated with
 Hawking radiation remain intact.

    Identical conclusions also hold for the gravitational case. Although
 not immediately obvious, a little algebra shows that the normal
 parity term in $\mathcal{A}_{t}$ (\ref{25}) vanishes. Hence the energy 
 momentum  flux (given by $T^{r}_{t}$) remains as before.\\

 {\it Discussions:-}\\
 This work was based on \cite{Isowilczek}but with
 a different procedure and emphasis. The flow of charge and energy momentum
 from charged black hole horizons were obtained by a cancellation of the
 covariant anomalies. Since the boundary condition involved the vanishing
 of the covariant current at the horizon, all calculations involved only
 covariant expressions.  Neither the consistent anomaly nor the counterterm
 relating the different currents, which were essential inputs in \cite{Isowilczek}, were required. Consequently our analysis was economical and, we feel,
 also conceptually clean. We would here like to mention that the interplay
 of covariant versus consistent anomalies, as occurring in 
\cite{Robwilczek,Isowilczek,Isoumtwilczek}, has been specifically discussed
 in the appendix of \cite{Morita}.  

    It should be pointed out that the flux is identified with $J^{r}_{(o)}$
 or $T^{r}_{t(o)}$ which are the expressions for the currents exterior
 to the horizon. Here these currents are anomaly free implying that
 there is no difference between the covariant and consistent expressions.
 Actually the germ of the anomaly lies in this difference\cite{rabin1,rabin2}.
 Hence it
 becomes essential, and not just desirable, to obtain the same flux in terms
 of the covariant expressions. In other words the Hawking flux must yield  
identical results
 whether one uses the consistent or the covariant anomalies. But
the boundary condition must be covariant.
This is consistent with the universality
 of the Hawking radiation and gives further credibility to the
 anomaly cancellation approach.

   It was shown \cite{Robwilczek, Isowilczek}, performing a partial wave decomposition, that 
physics near the horizon is descibed by an infinite collection of massless $(1+1)$ dimensional fields,
each partial wave propagating in spacetime with a metric given by the $`r- t'$ sector of the complete
spacetime metric (\ref{2}). This simplification, which effects a dimensional reduction from $d$-dimensions
to $d=2$ is also exploited here. It is however noted that greybody factors have not been included. In that 
case dimensional reduction will not yield the real Hawking radiation for $d>2$. For instance it is known
\cite{zelnikov} that for $d=4$ reduction to $d=2$ and keeping only the $s$-wave $(i.e. l=0)$ reduces the 
Hawking flux with respect to its $2d$ value. 

    A reason in favor of working with covariant anomalies is the fact
 that their functional forms are unique, being governed solely
 by the gauge (diffeomorphism) transformation properties. This is
 not so for consistent anomalies. They can and do have normal parity
 terms, apart from the odd parity ones. In fact, the special
 property (\ref{5}) of two dimensions yields a natural form for this
 anomaly which has normal parity terms. Our observation that the
 results of \cite{Robwilczek,Isowilczek} are still valid lend further support
 to this scheme of deriving Hawking radiation. The present approach can be
 easily extended to other (e.g rotating) black holes with Kerr-(Newman)
 metric.
    

     
\end{document}